\documentclass[letterpaper]{jpconf}
\usepackage{bm}%
\usepackage{graphicx}
\usepackage{epsfig}
\usepackage{amsmath,amssymb} 
\usepackage{booktabs} 
\usepackage{multirow} 
\usepackage[bookmarks]{hyperref} 

\input babarsym

\def\At   {\ensuremath{\mathcal{A}_T}\xspace}
\def\sys {\hbox{syst}}
\def\sta {\hbox{stat}}
\def\pisoftp    {\ensuremath{\pi_{\rm s}^{+}}\xspace}
\def\mKKpipi {\ensuremath{m(\Kp\Km\pip\pim)}\xspace}
\def\dm         {\ensuremath{\Delta m}\xspace}
\newcommand{\kevcc}{\ensuremath{{\mathrm{\,ke\kern -0.1em V\!/}c^2}}\xspace}

\begin{document}
\title{Search for \CP violation using \T-odd correlations in \Dz\to\Kp\Km\pip\pim decays at \babar.}

\author{Maurizio Martinelli (for the \babar\ Collaboration)}

\address{INFN and University of Bari, 70126 Bari, Italy.}

\ead{maurizio.martinelli@ba.infn.it}

\begin{abstract}
We search for \CP violation in a sample of $4.7\times10^4$ Cabibbo suppressed \Dz\to\Kp\Km\pip\pim decays.
We use 470\invfb of data
recorded by the \babar\  detector at the \pep2 asymmetric-energy \epem
storage rings running at center-of-mass energies near 10.6 GeV.
\CP violation is searched 
for in the difference between the \T-odd asymmetries, obtained using 
triple product correlations, measured for \Dz and \Dzb decays. 
The measured \At violation parameter is $\At = (1.0 \pm 5.1_{\sta}\pm 4.4_{\sys}) \times 10^{-3}$.
\end{abstract}

In Standard Model, \CP violation arises from Kobayashi-Maskawa phase in Cabibbo-Kobayashi-Maskawa quark mixing matrix\cite{Cabibbo:1963yz,Kobayashi:1973fv}. 
Theoretical attempts to predict the effect of \CP violation in Cabibbo suppressed charmed decays have been made in the past\cite{Buccella:1994nf}, obtaining a limit of 0.1\% not excluding even 1\% effects. 
The same paper suggests that this limit can be lowered by at least one order of magnitude by oscillations, which have been recently observed\cite{Aubert:2007wf,Abe:2007rd}.

\CP violation in charm decays can be exploited by many New Physics models\cite{Grossman:2006jg} both at tree and one-loop level; among these the latter expect a \CP violation asymmetry at the order $\mathcal{O}(10^{-2})$, which is now the level of experimental sensitivity\cite{Barberio:2008fa}.

We make use of \T-odd correlations\cite{Bigi:2001sg} to build a \T odd observable: assuming \CPT theorem, \CP violation is straightforward once \T violation is found.
One way to build a \T odd observable rely on the mixed product $\vec{v}_1\cdot(\vec{v}_2\times\vec{v}_3)$, where each $\vec{v}_i$ is a momentum or a spin.
A non-zero triple product correlation is then evidenced by the asymmetry
\[
A_T = \frac{\Gamma(\vec{v}_1\cdot(\vec{v}_2\times\vec{v}_3) > 0) - \Gamma(\vec{v}_1\cdot(\vec{v}_2\times\vec{v}_3) < 0)}
	{\Gamma(\vec{v}_1\cdot(\vec{v}_2\times\vec{v}_3) > 0) + \Gamma(\vec{v}_1\cdot(\vec{v}_2\times\vec{v}_3) < 0)},
\]
where $\Gamma$ is the decay rate of the process.
There is however a technical complication due to strong phases, which can fake this signal. The true \T violation observable is then
\[
\mathcal{A}_T = \frac{1}{2}(A_T - \bar{A}_T),
\]
where $\bar{A}_T$ is the charge conjugate of $A_T$, in which by definition the weak phase changes its sign, while the strong does not.
This observable can be built in the \Dz\to\Kp\Km\pip\pim decays defining $C_T \equiv \vec{p}_{\Kp}\cdot(\vec{p}_{\pip}\times\vec{p}_{\pim})$, using the momenta $\vec{p}_i$ of the final state particles in the \Dz rest frame, and taking ($\bar{C}_T \equiv \vec{p}_{\Km}\cdot(\vec{p}_{\pim}\times\vec{p}_{\pip})$)
\[
A_T = \frac{\Gamma(C_T > 0) - \Gamma(C_T < 0)}{\Gamma(C_T > 0) + \Gamma(C_T < 0)}\qquad
 \bar{A}_T = \frac{\Gamma(-\bar{C}_T > 0) - \Gamma(-\bar{C}_T < 0)}{\Gamma(-\bar{C}_T > 0) + \Gamma(-\bar{C}_T < 0)}.
\]

The reaction~\cite{conj} 
\[
\epem\to  X \ \Dstarp; \ \Dstarp\to \pip_s \Dz; \  \Dz \to  \Kp \Km \pip \pim,
\]
where $X$ indicates any system composed by charged and neutral particles, has been reconstructed from the sample of 
events having at least five charged tracks. We first reconstruct the $\Dz$ candidate: 
all $\Kp \Km \pip \pim$ combinations assembled from well-measured and 
positively identified kaons and pions are constrained to a common vertex.
To reconstruct the \Dstarp candidate,
we perform a vertex fit of the $\Dz$ candidates
with all combinations of charged tracks having a laboratory momentum below 0.65 \gevc $(\pisoftp)$ with the 
constraint that the
new vertex is located in the interaction region. 
We require the $\Dz$ to have a center-of-mass momentum greater than 2.5\gevc: this requirement removes $\Dz$ coming from $B$ decays.
We observe a contamination of the signal sample from $\Dz \to \Kp \Km \KS$, 
where $\KS \to \pip \pim$. The $\pip \pim$ effective mass shows, in fact, a distinct $\KS$ 
mass peak,
which can be represented by a Gaussian distribution with $\sigma=4.20 \pm 0.26$ $\mevcc$, which accounts for $5.2\%$ of the selected data sample. 
We veto $\KS$ candidates within a window of 2.5 $\sigma$.
This cut, while reducing to negligible level the background from \Dz~\to~\Kp~\Km~\KS, removes $5.8\%$ of the signal events. 

\begin{figure*}[htb]
\begin{center}
\includegraphics[width=\textwidth]{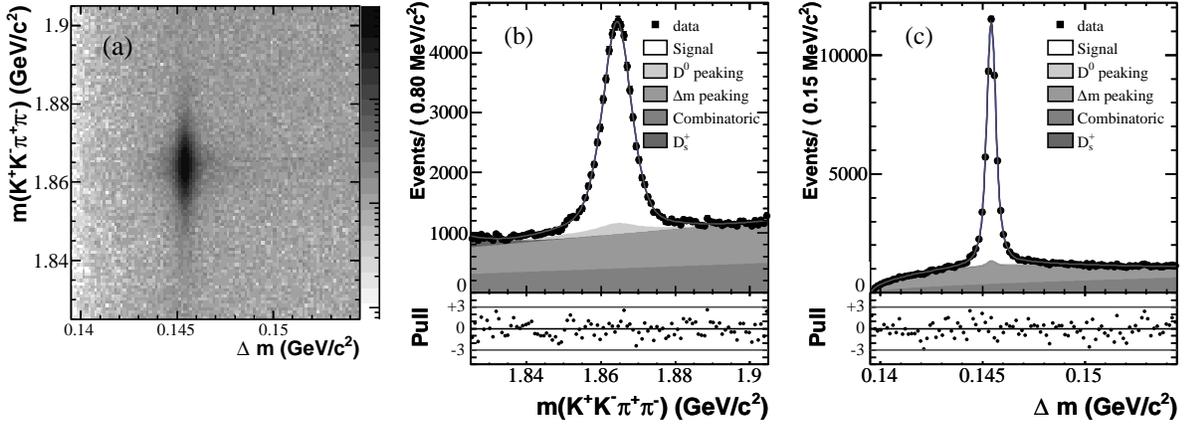}
\caption{(a) \mKKpipi vs. \dm for the total data sample. (b) \mKKpipi and (c) \dm projections with 
curves from the fit results.
Shaded areas indicate the different contributions. The fit residuals, represented by the pulls, are also shown under each distribution.}
\label{fig:fig1}
\end{center}
\end{figure*}

Defining the mass difference $\Delta m\equiv m(\Kp \Km \pip \pim \pisoftp) -  m(\Kp \Km \pip \pim)$,
Figure~\ref{fig:fig1}(a) shows the scatter plot $ m(\Kp \Km \pip \pim)$ vs. $\Delta m$  for all the events.
Figure~\ref{fig:fig1}(b) shows the $ m(\Kp \Km \pip \pim)$ projection, Fig.~\ref{fig:fig1}(c) shows 
the $\Delta m$ projection. 

We perform a fit to the \mKKpipi and \dm distributions,
using a polynomial background and a single Gaussian.
The fit gives $\sigma_{\Dz}=3.94 \pm 0.05\mevcc$ for the \Dz and 
$\sigma_{\Delta m}=244 \pm 20\kevcc$ 
for the $\Delta m$. 
We define the signal region within 
$\pm 2  \sigma_{\Dz}$ and $\pm 3.5  \sigma_{\Delta m}$.
The total yield of tagged \Dz mesons in the signal region is approximately $4.7 \times 10^4$ events.


The \Dz yields to be used in the calculation of the \T asymmetry
are determined using a binned, extended
maximum-likelihood fit to the 2-D (\mKKpipi, \dm) distribution obtained with the two
observables \mKKpipi and \dm in the mass regions defined in the ranges $1.825<\mKKpipi<1.915 \ \gevcc$ and $0.1395<\dm<0.1545 \ \gevcc$ respectively.
Events having more than
one slow pion candidate in this mass region are removed (1.8 \% of the final sample).
The final 2-D distribution contains approximately $1.5 \times 10^5$ events
and is divided into a $100 \times 100$ grid.  
 
The 2-D (\mKKpipi, \dm) distribution is described by five components:
\begin{enumerate}
\item True \Dz signal originating from a \Dstarp decay. This component has characteristic peaks in both observables \mKKpipi and \dm.
\item Random \pisoftp events where a true \Dz is associated to an incorrect \pisoftp, called \Dz peaking.
This contribution has the same shape
in \mKKpipi as signal events, but does not peak in \dm.
\item Misreconstructed \Dz decays where one or more of the \Dz decay products are either not
reconstructed or reconstructed with the wrong particle
hypothesis, called \dm peaking. Some of these events show a peak in \dm, but not in \mKKpipi.
\item Combinatorial background where the \Kp, \Km, \pip, \pim candidates are not fragments of the same \Dz decay, called combinatoric.  
This contribution does not exhibit any peaking structure in
\mKKpipi or \dm. 
\item \Ds\to\Kp\Km\pip\pim\pip contamination, called \Ds. 
This background has been studied on Monte Carlo (MC) simulations and shows a characteristic linear narrow shape in the 2-D (\mKKpipi, \dm) distribution,
too small to be directly visible in Fig.~\ref{fig:fig1}(a).
\end{enumerate}

The functional forms of the probability density functions (PDFs) for
the signal and background components are based on studies of
MC samples. 
However, all parameters related to these functions are determined from
two-dimensional likelihood fits to data over the full \mKKpipi vs.\ \dm
region. We make use of combinations of Gaussian and Johnson SU~\cite{jsu} lineshapes for peaking distributions, and we use polynomials and
threshold functions for the non-peaking backgrounds. 
The results of the fit are shown in Fig.~\ref{fig:fig1}. The fit residuals shown under each distribution are represented by $Pull=(N_{data} - N_{fit})/\sqrt{N_{data}}$.


According to the \Dstarp tag and the $C_T$ variable, we divide the total data sample into four subsamples, defined in Table~\ref{tab:ct}.
These four data samples are fit simultaneously to the same model.
The signal event yields are given in Table~\ref{tab:ct}.

\begin{table}[!htb]
\caption{Definition of the four subsamples and the event yields from the fit.}
\begin{center}
\begin{tabular}{c|c}
\hline
Subsample & Events \cr
\hline
(a) \Dz, $C_T>0$ &  10974 $\pm$ 117 \cr
(b) \Dz, $C_T<0$ & 12587 $\pm$ 125  \cr
(c) \Dzb, $\overline {C}_T>0$ & 10749 $\pm$ 116  \cr
(d) \Dzb, $\overline {C}_T<0$ &  12380 $\pm$ 124 \cr
\hline
\end{tabular}
\label{tab:ct}
\end{center}
\end{table}


We validate the method using \epem \to \ccbar MC simulations, where \Dz  
decays through the intermediate resonances with the branching  
fractions reported in the PDG~\cite{pdg}. We obtain a \T asymmetry 
$\At=(2.3\pm 3.3)\times 10^{-3}$, consistent with the generated  
value of $1.0\times 10^{-3}$.

To test the effect of possible asymmetries generated by the detector, we use signal MC 
in which the \Dz decays uniformly over phase space.
In this case possible asymmetries are generated only by the detector efficiency:
$\At= -(1.1 \pm 1.1) \times 10^{-3}$,
again consistent with zero.

To avoid potential bias, all event selection criteria are determined
before evaluating $\mathcal{A}_T$.
Systematic uncertainties are obtained directly from the data. In these studies
the true $A_T$ and $\overline{A}_T$ central values are masked by adding unknown random offsets.
Removing the offsets:
\begin{equation}
A_T = (- 68.5 \pm 7.3_{\sta} \pm 5.8_{\sys}) \times 10^{-3}\qquad
\overline{A}_T =  (- 70.5 \pm 7.3_{\sta} \pm 3.9_{\sys}) \times 10^{-3}.
\end{equation}

We observe non-zero values of $A_T$ and $\overline{A}_T$ indicating that final state interaction
effects are significant in this \Dz decay. No effect is found, on the other hand, in the analysis of MC samples.

The result for the \CP violation parameter, \At, is
\begin{equation}
\At = (1.0 \pm 5.1_{\sta}\pm 4.4_{\sys}) \times 10^{-3}. 
\end{equation}


The sources of systematic uncertainties considered in this analysis 
and
the estimates of their values are derived as follows:
\begin{enumerate}
\item The PDFs used to describe the signal are modified, replacing the Johnson SU function by a Crystal Ball
function~\cite{cb}, obtaining fits of similar quality ($\sigma_{sys} = 0.2\times10^{-3}$).
\item As the same as (i),  for the peaking background ($\sigma_{sys} = 0.5\times10^{-3}$).
\item We increase the number of bins of the 2-D (\mKKpipi, \dm) distribution to a ($120 \times 120)$ grid and decrease to a grid of $(80 \times 80)$ ($\sigma_{sys} = 0.2\times10^{-3}$). 
\item The particle identification algorithms used to identify kaons and pions are modified to more 
stringent conditions in different combinations ($\sigma_{sys} = 3.5\times10^{-3}$). 
\item The $p^*(\Dz)$ cut is increased to 2.6\gevc and 2.7\gevc ($\sigma_{sys} = 1.7\times10^{-3}$).
\item We study possible intrinsic asymmetries due to the interference
between the electromagnetic $e^+e^-\to \gamma^* \to \ c\bar{c}$ and weak
neutral current $e^+e^-\to Z^0\to \ c\bar{c}$ amplitudes. 
This interference produces a \Dz/\Dzb production asymmetry
that varies linearly with the quark production angle with respect to the
$e^-$ direction. We constrain the possible 
systematics by measuring $\mathcal{A}_T$ in three regions of 
the center-of-mass \Dz production angle $\theta^*$: forward ($0.3 < \cos(\theta^*)_{\Dz}$), 
central ($-0.3 < \cos(\theta^*)_{\Dz} \leq 0.3$), and backward ($ \cos(\theta^*)_{\Dz} < -0.3$) ($\sigma_{sys} = 0.9\times10^{-3}$).
\item Fit bias: we use MC simulations to compute the difference between the generated and reconstructed ${\cal A}_T$ ($\sigma_{sys} = 1.4\times10^{-3}$).
\item Mistag: there are a few ambiguous cases with more than one $D^*$ in the event. We use MC simulations where
these events are included or excluded from the analysis. This effect has a negligible contribution to the systematic
uncertainty. 
\item Detector asymmetry: we use the value obtained from the MC simulation where \Dz decays unifromly 
over the phase space ($\sigma_{sys} = 1.1\times10^{-3}$).  
\end{enumerate}
In the evaluation of the systematic uncertainties, we keep, for a given category, 
the largest deviation from the reference value
and assume symmetric uncertainties. Thus, most systematic uncertainties are statistical
in nature, and are conservatively estimated.



In conclusion, we search for \CP violation using $T$-odd correlations in a high statistics sample of Cabibbo suppressed 
$\Dz \to \Kp \Km \pip \pim$ decays. We obtain a $T$-violating asymmetry consistent with zero with a 
sensitivity of $\approx$ 0.5 \%. 
These results constrain the possible effects of New Physics in this observable.


\section*{References}
\begin{thebibliography}{9}

\bibitem{Cabibbo:1963yz}
  N.~Cabibbo,
  Phys.\ Rev.\ Lett.\  {\bf 10}, 531 (1963).

\bibitem{Kobayashi:1973fv}
  M.~Kobayashi and T.~Maskawa,
  Prog.\ Theor.\ Phys.\  {\bf 49} (1973) 652.
  
\bibitem{Buccella:1994nf}	
  F.~Buccella, M.~Lusignoli, G.~Miele, A.~Pugliese and P.~Santorelli,
  Phys.\ Rev.\  D {\bf 51}, 3478 (1995)
    
\bibitem{Aubert:2007wf}
  B.~Aubert {\it et al.}  [BABAR Collaboration],
  Phys.\ Rev.\ Lett.\  {\bf 98} (2007) 211802.

\bibitem{Abe:2007rd}
   L.~M.~Zhang {\it et al.}  [BELLE Collaboration],
	Phys. Rev. Lett. {\bf 99}, 131803 (2007).

\bibitem{Grossman:2006jg}
  Y.~Grossman, A.~L.~Kagan and Y.~Nir,
  Phys.\ Rev.\  D {\bf 75}, 036008 (2007).

\bibitem{Barberio:2008fa}
  E.~Barberio {\it et al.}  [\href{http://www.slac.stanford.edu/xorg/hfag/index.html}{Heavy Flavor Averaging Group}],
  arXiv:0808.1297 [hep-ex].

\bibitem{Bigi:2001sg}
  I.~I.~Y.~Bigi,
  arXiv:\hepph{0107102}.
  
\bibitem{conj}
Charge-conjugate reactions are implied throughout.

\bibitem{geant}
S. Agostinelli {\it et al.} (\textsc{Geant4} Collaboration), Nucl. Instrum. Methods
Phys. Res., Sect A {\bf 506}, 250 (2003).

\bibitem{jsu}
N.L. Johnson, Biometrika {\bf 36}, 149 (1949).

\bibitem{cb}
J.E. Gaiser, Appendix-F 
of Ph.D. Thesis, SLAC-R-255 (1982).

\bibitem{pdg}
C. Amsler {\it et al.} (Review of Particle Physics), Phys. Lett. {\bf B667}, 1 (2008).

\bibitem{asy_babar} B. Aubert {\it et al.} (\babar \ Collaboration), Phys. Rev. Lett. {\bf 100}, 061803 (2008).

\end{thebibliography}
\end{document}